\documentclass[aps,prl,preprint,groupedaddress]{revtex4}
\usepackage[toc,page]{appendix}
\usepackage{graphicx,pdflscape,epsf,epsfig,amsmath}% Include figure files
\usepackage{alltt,dsfont,amsmath,bm}

\newcommand{\llabel}[1]{\label{#1} }

\newcommand{\iden}{ \mathds{ 1}}

\newcommand{\up}{\Upsilon}

\newcommand{\G}{{\cal{G}}}

\newcommand{\GH}{{\bf g}}
\newcommand{\GHI}{\GH^{-1}}

\newcommand{\LL}{{\bf L}}

\newcommand{\si}{\sigma}
\newcommand{\sib}{\bar{\sigma}}

\newcommand{\tJ}{\ $t$-$J$ \ }

\newcommand{\U}{{\cal U}}

\newcommand{\V}{{\cal V}}
\newcommand{\I}{{\bf I}}

\newcommand{\bb}[1]{{\bar{\mathbf #1}}}
\newcommand{\eb}{\overline{\varepsilon}}

\newcommand{\nn}{\nonumber}
\newcommand{\chem}{{\bm \mu}}

\newcommand{\beq}{\begin{equation}}
\newcommand{\eeq}{\end{equation}}
\newcommand{\barray}{\begin{eqnarray}}
\newcommand{\earray}{\end{eqnarray}}
\newcommand{\disp}[1]{Eq.~(\ref{#1})}
\newcommand{\refdisp}[1]{Ref.~(\onlinecite{#1})}
\newcommand{\figdisp}[1]{Fig.~(\ref{#1})}

\begin{document}
\title{Extremely Correlated Fermi Liquids}
\author{ B. Sriram Shastry}
\affiliation{ Physics Department, University of California, Santa Cruz, CA 95064, USA}
\date{\today}

\begin{abstract}
We present the theory of an extremely correlated  Fermi liquid with $U\to \infty$. This liquid has an underlying auxiliary Fermi liquid  Greens function that is further caparisoned by extreme correlations. The theory  leads to two parallel  hierarchies of equations that permit iterative approximations in a certain parameter. Preliminary results for the spectral functions display a broad background and a distinct $T$ dependent left skew. An important energy scale $\Delta(\vec{k},x)$ emerges as the average inelasticity of the FL Greens function, and influences the photoemission spectra profoundly. A duality is identified wherein  a loss of coherence of the ECFL  results from an excessively   sharp FL.
\end{abstract}

%\pacs{72.10.Bg, 72.15.Gd, 75.47.-m, 72.80.Ng, 71.70.Ej}

\maketitle

\date{\today}
\maketitle

{\bf Introduction}
Correlated electron systems attract two distinct approaches. An intermediate to strong coupling approach is used when the interaction $U$ is comparable to the band width $2 W$, and has seen some success in recent times\cite{dmft}. On the other hand, Anderson\cite{anderson} has   argued  that myriad experiments on high $T_c$ superconductors require a better understanding of the \tJ model physics. This model sets $U \to \infty$  right away i.e. leads to   {\em extreme correlations}  and involves  Gutzwiller projected Fermi operators that are { non canonical}.  Thus Wick's theorem  is immediately lost, and perturbative schemes encoding  the Feynman Dyson approach become useless. Since this approach is at the root of most current many body physics text books,  the task of   understanding  the \tJ model is not lightly undertaken. 

 The  Schwinger approach to interacting field theories  is a powerful and  attractive  alternative.
  It is fundamentally  non perturbative,  where  Wick's theorem is bypassed by dealing with suitable inverse Greens functions. Conventional many body theory for canonical Fermions can also be  cast into this  approach, and leads to the standard  results.  In Ref.  \onlinecite{ECQL} (henceforth \I), the author has recently applied the Schwinger method to the \tJ model, and found a class of solutions that are termed as extremely correlated quantum liquids. That state  is presumably realized under  suitable conditions. However it gives a Fermi surface  (FS) volume that is always distinct from that of the Fermi gas. This is  contrast to the case of Fermi liquids (FL), where the important theorem of  Luttinger and  Ward  (L-W) \cite{luttinger,agd} mandates the invariance  of the FS volume under interactions.

In this paper    we propose a state of matter termed as an   {\em extremely correlated Fermi  liquid} (ECFL). The ECFL found here, represents an   alternate class of solutions for the \tJ model, where the Fermi surface  satisfies   the Fermi gas (i.e. L-W)  volume.
%added
 In this work we present the essentials of the formalism,  and display  preliminary results on spectral functions that are suggestive of the relevance of the ECFL  state to cuprate materials.
%end added
   An inherent flexibility of the  Schwinger approach   permits the construction of an alternate class of solutions from the one found in \I.   The excitations of the ECFL state  may be thought of as  bare electrons undergoing a double layer of renormalization: the FL dressing into  quasiparticles that are further    caparisoned (i.e. decorated)  by extreme correlations. 

{\bf Formalism:}
The physical projected electronic Greens function $\G$ satisfies an equation of motion (EOM)  (\I-29) written   compactly in matrix form as
\barray
&& (\partial_{\tau_i} - \chem) \G(i,f)  =  - \delta(i,f)  \left\{ 1- \gamma(i) \right\} - \V_i \cdot \G(i,f) \nn \\
&& - X(i,\bb{j}) \cdot \G(\bb{j},f) -Y(i,\bb{j}) \cdot \G(\bb{j},f) \llabel{eom_2},
\earray  
where $\chem$ is the chemical potential and  an implicit  integration over space time variables such as $\bb{j}$, written with  bold  overlined letters,    is implied, 
\barray
X(i,j)&=& - t(i,j)  \  (D(i)+D(j))+ \frac{1}{2} J(i,\bb{k}) \  (D(i)+ D(\bb{k})) \delta(i,j) \nn \\
Y(i,j)&=& - t(i,j)  \  (\iden- \gamma(i)- \gamma(j))+ \frac{1}{2} J(i,\bb{k}) \  ( \iden - \gamma(i)- \gamma(\bb{k})) \delta(i,j). 
\earray
In the above expression\cite{footnote-sym}, we used
$
\gamma({i}) =  \G^k(i,i) 
$
with the $k$ conjugation defined by $(M^k)_{\si_1 \si_2}=M_{\sib_2 \sib_1} \si_1 \si_2 $, and $D_{\si_1 \si_2}(i)= \si_1 \si_2 \frac{\delta}{\delta \V_i^{\sib_1 \sib_2 }}$ . 
%added
The added (Bosonic)  source term $\V_i^{\si_1 \si_2}(\tau_i)$ is central to this approach; it  is a space-time   dependent field  that couples to the charge and spin densities through a term in the action: $\sum_{i \si} \int_0^\beta d \tau \ \V_i^{\si_1 \si_2}(\tau) \ X^{\si_1 \si_2}_i(\tau) $, where $X_i^{\si_1 \si_2}$ is the  spin and  density operator at site $i$ that acts as $|\si_1 \rangle \langle \si_2 |$ .
%end added

An important technical  problem highlighted in \I ~    is to deal with the time dependence of the  $\gamma(i)$ term in \disp{eom_2} which makes the theory non canonical. Here we use  the decomposition into two factors \cite{fn1} :
\barray
\G(a,b)&=& \GH(a,\bb{b}) \cdot \mu(\bb{b},b), \label{decomposition}
\earray
and express  $\gamma(i)= (\GH(i,\bb{j})\cdot \mu(\bb{j},i))^k$.
 The object $\GH$  is  an auxiliary  FL   Greens function and $\mu(\bb{b},b)$ is  an   appurtenant (or supplementary)  factor   that  is determined below. 
 %added
 Antiperiodic boundary conditions  $\G(0,\tau_f)=-\G(\beta,\tau_f)$ and $ \G(\tau_i,0)=-\G(\tau_i,\beta)$ imply that   both  factors $\GH$ and $\mu$  are Fourier transformed using Fermionic Matsubara frequencies.
 %end added
 We define the inverse Greens function $\GHI(a,\bb{b}) \cdot \GH(\bb{b},b)= \iden \delta(a,b)$, and thence  a vertex function
$
\Lambda^{\si_1 \si_2}_{\si_3 \si_4}(p,q;r) =   - \frac{\delta}{\delta \V^{\si_3 \si_4}_r} \ \{ \GHI_{\si_1 \si_2}(p,q) \ \}.
$ 
  Thus $\GH$, $\mu$  and $\GHI$   are matrices in the spin space, and the vertex $\Lambda$ has four indices. 
We  also define a linear operator
\beq
\LL(i,f)  =  \left( t(i,\bb{j}) \ \xi^* \cdot \GH(\bb{j},f) - \frac{1}{2} J(i,\bb{j}) \  \xi^* \cdot \GH(i,f)\right)\cdot \left( \frac{\delta}{\delta \V_i^*} +  \frac{\delta}{\delta \V_\bb{j}^*} \right),  \label{eq-L}
\eeq
where the matrix  $\xi^*_{\si_1 \si_2}=\si_1 \si_2$. The $^*$ is used as a place holder that transmits the spin indices (after conjugation) of the $\xi$ matrix to the source matrix $\V$ in the functional derivative. This  notation used is illustrated  in component form by 
$
\cdots \xi^*_{\si_a \si_b}\cdots  {\delta}/{\delta \V_\bb{j}^*}= \cdots \si_a \si_b \cdots \delta/{\delta \V_\bb{j}^{\sib_a, \sib_b }}.
$

A useful chain rule for the functional derivative is noted
\barray
D(r) \cdot \G(a,b)& = & \xi^* \cdot \GH(a,\bb{c}) \cdot \  \Lambda_*(\bb{c},\bb{d};r) \ \cdot \G(\bb{d},b) 
\nn \\
&&+ \xi^* \cdot  \GH(a,\bb{b})   \cdot \ \left( \frac{\delta}{\delta \V^{*}_r} \mu(\bb{b},{b})\right)
\earray
Using this chain rule, we see that 
\barray
X(i,\bb{j}) \cdot \G(\bb{j},f) &\equiv& \Phi(i,\bb{b})\cdot \G(\bb{b},f) + \Psi(i,f) 
\earray
where
\barray
 \Phi(i,m)= \LL(i,\bb{i}) \cdot \GHI(\bb{i},m)  \nn \\
 \Psi(i,m) = - \ \LL(i,\bb{i}) \cdot \mu(\bb{i},m) \label{phi-psi-1}
 \earray
Thus the two fundamental functions of this formalism $\Phi, \Psi$ are closely connected as they arise from applying the same operator to the two factors of $\G$. 
Defining
$
Y_0(i,j)= \left( -t(i,j)+  \frac{1}{2} J(i,\bb{k}) \  \delta(i,j)\right) \iden $, and
 $
Y_1(i,j)= t(i,j)  \  (\gamma(i)+\gamma(j)) -  \frac{1}{2}\ \delta(i,j) \  J(i,\bb{k}) \  (\gamma(i)+ \gamma( \bb{k})) $,
  also denote the Fermi gas Greens function
\beq
\GHI_0(i,f)=
 \{- (\partial_{\tau_i} - \chem)\iden - \V_i)\delta(i,f) - \ Y_0(i,f)  \}.
\eeq
 Collecting everything,  the exact  EOM can now be written neatly as
\barray
&& \{ \GHI_0(i,\bb{j}) - \lambda  \ Y_1(i,\bb{j}) - \lambda \ \Phi(i,\bb{j}) \} \ \cdot \GH(\bb{j},\bb{f})\cdot \mu(\bb{f},f)   \nn \\
&&=   \delta(i,f)  \left( \iden - \lambda \ \gamma(i)  \right)   + \lambda \ \Psi(i,f). \label{EOM-1}
\earray 
We have introduced  the parameter $\lambda$ above, with $0 \leq \lambda \leq 1$, in order to provide an adiabatic  path between the Fermi gas at $\lambda=0$ and the ECFL at $\lambda=1$, and also  an iterative scheme in powers of $\lambda$   connecting the two endpoints.

We  now  choose the hitherto undetermined function $\mu$ as:
\beq
\mu(i,f)   =   \delta(i,f)   \left( \iden - \lambda \ \gamma(i) \right)     + \lambda \ \Psi(i,f), \label{EOM-2}
\eeq 
so that    \disp{EOM-1} reduces to a canonical FL type equation:
\beq
\{ \GHI_0(i,\bb{j}) - \lambda  \ Y_1(i,\bb{j}) - \lambda \  \Phi(i,\bb{j}) \} \ \cdot \GH(\bb{j},f)   =   \delta(i,f). \label{EOM-3}   
\eeq
Notice that    the right hand side has a pure  $\delta$  function  as in a canonical Fermi liquid type theory. To summarize, the EOM \disp{eom_2}  under  the decomposition \disp{decomposition} leads to \disp{EOM-1}.  In turn this  splits exactly  into two coupled  sets of equations \disp{phi-psi-1}, \disp{EOM-2} and \disp{EOM-3} for the two factors $\GH$ and $\mu$. Note that the entire procedure is exact,  we write explicit forms of these equations   below and then introduce approximate methods to solve them .

Inverting  we find  Dyson's equation for the auxiliary  FL Greens function:
\beq
 \GHI(i,m)     =  \{\GHI_0(i,m) - \lambda \ Y_1(i,m) - \lambda \ \Phi(i,m) \}.   \label{EOM-4}
 \eeq
Taking functional derivatives of \disp{EOM-2} and \disp{EOM-4} w.r.t. $\V$,  and comparing with  \disp{eq-L} and \disp{phi-psi-1} we generate 
{\em two parallel hierarchies of  equations} for $\GH$  and $\mu$ that form the core of this formalism. 
The hierarchy for $\GH$ is essentially autonomous and drives  that for $\mu$. 
Starting with the Fermi gas at $O(\lambda^0)$, an iterative process similar to the skeleton graph expansion of L-W\cite{luttinger} can be built up, such that  terms of $O(\lambda^n)$ arise from differentiating lower order terms of  $O(\lambda^{n-1})$. Systematic approximations may thus be arranged to include all terms of $O(\lambda^n)$ for various $n$\cite{lambda}. 
% added
The number of particles is given by
$
\frac{1}{2} n(i)= \GH(i,\bb{i}) \cdot \mu(\bb{i},i),  
$
and with
\beq\U^{\si_1 \si_2}_{\si_3 \si_4}(a,b;c) \equiv  \frac{\delta \mu_{\si_1 \si_2}(a,b)}{\delta \V_c^{\si_3 \si_4}},
\eeq
the equations to solve simultaneously are \disp{phi-psi-1},  \disp{EOM-4} and  \disp{EOM-2}. The density and spin density  response functions (\I-F1,\I-F-7) can be found from differentiating $\G$ i.e.
$ \up^{\si_1 \si_2}_{\si_3 \si_4}(p,q;r) =  \frac{\delta }{\delta \V_c^{\si_3 \si_4}} \ \left\{  \G_{\si_1 \si_2}(p,q)\right\}.  $

{\bf Zero source limit in  Fourier space:}
When we turn off the source $\V$, the various matrix function $\G, \GH, \mu$ become spin diagonal and translation invariant so we can Fourier transform these conveniently.
We note the basic result expressing $\G$ as a simple product of two functions in $k$ space:
\barray
\G(k)& =& \ \GH(k) \ {\mu(k)}, 
\;\;\; \mu(k)  =  1- \lambda \  \frac{n}{2}   + \lambda \  \Psi(k) \nn \\
\GHI(k) &= &  i \omega_k + \chem - \varepsilon_k (1 - \lambda \ n)  - \lambda \ \Phi(k) \label{g-1}
\earray
where $\varepsilon_k$ is the Fourier transform of the hopping matrix $-t(i,j)$, and an uninteresting   constant term  is absorbed in $\chem$ here and below.

 Here $\GH$ plays the role of an underlying auxiliary  FL with a self energy $\Phi$, and $\Psi$ acts as an extra spectral weight that vanishes at high frequency,  leaving the exact weight $1- \frac{n}{2}$ valid for a projected electron (as in \I) for  $\lambda=1$.
 Denoting $\sum_{k}\to \frac{1}{N_s \ \beta}\sum_{ i \omega_k, \vec{k}}$ with $N_s$ sites, 
the particle number sum rule is $\sum_k \mu(k) \GH(k) =\frac{n}{2}$, i.e.
\beq
\frac{n}{2} = \sum_k  \GH(k) +\lambda  \sum_k ( \Psi(k) - \frac{n}{2} ) \ \GH(k). \label{number-sumrule}
\eeq
In this formalism, at $k\sim k_F, \ x=0$ that is relevant to the L-W sum rule,  the $\Re e \ \GH(k)$ dominates  $\Re e \ \G(\vec{k},0)$ (since $\Re e \ \Psi(\vec{k},0)$ is smooth through the FS). Requiring consistency with  the L-W  theorem  forces us to  pin any sign change  of $\Re e \ \GH(\vec{k},0)$ to the free case, whereby  we  impose a {\em second level sum rule}
\beq
 \sum_k  \Psi(k)  \ \GH(k)=  \frac{n^2}{4}, \;\;\mbox{and}\;\; \sum_k     \GH(k)=  \frac{n}{2}.\label{higher-order-sumrule}
\eeq  
This can be viewed as a splitting of the usual number sum rule \disp{number-sumrule} \cite{fn2}.
With
$
E(p_1,p_2)= \left( \varepsilon_{p_1}+\varepsilon_{p_2} +\frac{1}{2} \hat{J}(0)+ \frac{1}{2} \hat{J}(p_1-p_2) \right)$ we find
\barray
 \Phi(k)&=&  \ \sum_p  \ E(k,p) \ \GH(p) \ \Lambda^{(a)}(p,k)\nn \\
\Psi(k)&=&  \ \sum_p  \ E(k,p) \  \GH(p) \
   \U^{(a)}(p,k) 
\earray
and  the spin labels are from \I ~ with the usual significance $\Lambda^{(a)}= \Lambda^{(2)} -\Lambda^{(3)} = \frac{1}{2} \Lambda^{(s)} -  \frac{3}{2} \Lambda^{(t)} $.

Next we introduce the spectral representation of various functions $Q$ that vanish at infinity:
$
Q(i  \omega_Q) =  \int_{-\infty}^\infty \ dx \ \frac{\rho_Q(x)}{i  \omega_Q - x}
$
and
$
\rho_Q(x)  =  - \frac{1}{\pi} \ \Im m \ Q(x + i 0^+),
$ with $x^+\equiv x+ i 0^+$. The Matsubara frequency $\omega_Q$ is Fermionic  (Bosonic) if $Q$ is Fermionic (Bosonic).  Proceeding further, at any order in $\lambda$,  the two hierarchies give us coupled equations for the spectral densities of the physical particles  $\rho_{\G}(\vec{k}, x)$ as well as   the underlying Fermi liquid  $\rho_{\GH}(\vec{k},x)$,  in terms of the two  objects $\rho_{\bar{\Phi}}(\vec{k},x)$  and $\rho_\Psi(\vec{k},x)$ and their Hilbert transforms.   The  Lehmann representation implies that $ \rho_{\G}(\vec{k}, x) $ is positive at all $\vec{k},x$. In making approximations, this important and challenging constraint must be kept in mind.  

%The density of particles is given by using \disp{higher-order-sumrule}
%$
%\frac{n}{2}=   \int_{-\infty}^\infty \ dx \ f(x) \ \sum_{\vec{k}} \rho_{\G}(\vec{k}, x). 
%$

 {\bf Solution of $\GHI$ and $\mu$ to   order $O(\lambda)^2$:}
 We next discuss a systematic expansion in powers of $\lambda$ \cite{lambda}, obtained by taking functional derivatives of \disp{EOM-2} and \disp{EOM-4} to generate expressions for the vertices given the Greens functions via $\Lambda\sim - \frac{\delta}{\delta \V} \GHI$ and 
 $\U \sim  \frac{\delta}{\delta \V} \mu$.  
 To  lowest order  in $\lambda$ ,   the bare vertex  $\Lambda^{(a)}=-1$,   this term  is absorbed in  a renormalization of the band dispersion to $\eb_k$ in \disp{g-1} \cite{dispersion-renorm} , and the remaining term denoted by $\bar{\Phi}(k)$. To this order $\U^{(a)}=0$. Proceeding to the next   non trivial order in $\lambda$, by taking the functional derivative of \disp{EOM-2} and \disp{EOM-4} we find after a brief calculation:
\barray
\Psi(k) &=&  -2  \lambda \ \sum_{p,q}  \ E(k,p) \  \GH(p) \ \GH(q) \ \GH(q+p-k)   \nn \\
\bar{\Phi}(k)&=& - 2 \lambda \ \sum_{p,q}  \ E(k,p) \left(E(p,k)+E(q+p-k,p) \right) \nn \\
&&  \GH(p) \ \GH(q) \ \GH(q+p-k). \label{order-lambda}
\earray
From \disp{g-1} we note that these expressions \disp{order-lambda} lead to a   calculation of  $\GHI$  and $\mu$  correct upto $O(\lambda^2)$.  Frequency dependent corrections arise only to second order in $\lambda$, which is  analogous to  the  structure  of the canonical many body theory  within the skeleton graph expansion. 
We may now set $\lambda =1$ and study the resulting theory as the first step in exploring this formalism.

Denote $f(x)= \frac{1}{(\exp{\beta x})+1}$ as the  Fermi distribution functions and  $\bar{f}(x)=1-f(x) $ , and denote the usual Fermi factors from second order theory 
$${\cal W}= \left\{f(u)f(w)\bar{f}(v)+f(v)\bar{f}(u)\bar{f}(w) \right\}  \delta(u+ w-v - x), $$
a function of the frequencies $u,v,w,x$, and
\beq{\cal Y}= \ \int_{u,v,w} {\cal W} \  \rho_{\GH}(\vec{q},w)  \rho_{\GH}(\vec{p},u)  \rho_{\GH}(\vec{q}+ \vec{p}-\vec{k},v), \label{Y}
\eeq
a function of $\vec{k},\vec{p},\vec{q}$ and $x$. We  may then write the spectral functions corresponding to  \disp{order-lambda}
\barray
\rho_{\bar{\Phi}}(\vec{k},x)&=& 2   \sum_{\vec{p},\vec{q}}   E(\vec{k},\vec{p})  \left(E(\vec{p},\vec{k})+E(\vec{q}+\vec{p}-\vec{k},\vec{p}) \right)  {\cal Y}  \nn \\
\rho_{\Psi}(\vec{k},x)&=&  2  \ \sum_{\vec{p},\vec{q}}  \ E(\vec{k},\vec{p})\ {\cal Y}.
 \label{eqs-phi-psi-1}
\earray
The functions appearing in \disp{eqs-phi-psi-1} are familiar from 
Fermi liquids\cite{agd,luttinger}, and encode the usual phase space constraints of that theory. This leads to the low temperatures  behaviour 
$\sim \mbox{max} \left\{ x^2, (\pi k_BT)^2 \right\}$,
for both objects  $\Im m \ \Psi(k,x,T)$ and  $\Im m \ \bar{\Phi}(k,x,T) $. The  real parts  of these objects are smooth through the Fermi surface, as one expects from the real part of the self energy in a FL, and hence motivates the second level sum rule \disp{higher-order-sumrule}.

From \disp{g-1} we write the exact expression for the physical spectral function $\rho_{\G}$:     
\beq
\rho_{\G}(\vec{k}, x) = \rho_{\GH}(\vec{k},x) \left(   \left\{ 1- \frac{n}{2} \right\} \ +
 \frac{\xi_k - x}{\Delta(\vec{k},x)}+\eta(\vec{k},x)  \right), \label{spectral-2}
\eeq
where   $\xi_k= \hat{\varepsilon}_k-\chem $, and  the important  energy scale $\Delta(\vec{k},x) $ and the term $\eta$  is defined as:
\barray
\Delta(\vec{k},x) &=& -\frac{\rho_{\bar\Phi}(\vec{k},x)}{\rho_\Psi(\vec{k},x)},  \label{delta} \\
\eta(\vec{k},x) & =&  \Re e  \Psi(\vec{k},x^+) +  \frac{1}{\Delta(\vec{k},x)}  \Re e  \Phi(\vec{k},x^+). ~
\label{eta}
\earray
The sign of the energy scale $\Delta$  in \disp{delta}  is expected to be positive   from \disp{eqs-phi-psi-1}. 
The dimensionless term $\eta$ augments the spectral weight at the Fermi level.
The equations necessary to solve the theory to $O(\lambda^2)$  may be summarized as 
\disp{g-1}, \disp{higher-order-sumrule}, \disp{order-lambda} and \refdisp{dispersion-renorm}
giving rise to the spectral function \disp{spectral-2}. 
These require further numerical work that is underway, it leads to spectral functions in 2 and 3 dimensions  that will be published separately. However it  also provides a very interesting insight  about the theory in high dimensions that is  pursued analytically next.

{\bf Solution in high dimensions:}
 In sufficiently high  dimensions, we show next that the dimensionless term $\eta$  vanishes identically leading to a great simplification.   For sufficiently high dimensions we can    ignore the momentum dependence of ${\cal Y}$ in \disp{Y} and assume  $\rho_{\Phi}(\vec{k},x) \sim C_\Phi \ \sigma(x)$,  and   $\rho_{\bar{\Psi}}(\vec{k},x) \sim C_\Psi \ \sigma(x)$,   as functions of frequency only. Here  $\sigma(x)$  extends over energy range $\omega_c \sim O(2 W)$, and   $C_\Phi$   has dimensions of inverse energy and is positive  due to  $\rho_{\bar{\Phi}}$.   Its Hilbert transform is called $ h(x)\equiv {\cal P} \ \int \ dy \ \frac{\sigma(y)}{x-y} $.  
 We use an  analytically tractable  Fermi liquid  model\cite{fn6}  with $\tau= \pi k_B T$, where    we set:
\beq
\sigma(x) =  \  \{ x^2+  \tau ^2  \}  e^{- C_\Phi \{ x^2+ \tau^2 \}/\omega_c }\ . \label{model-sigma} 
\eeq
 The peak value of $C_\Phi \  \sigma(x) $ is of $O(1)$ and independent of $C_\Phi$ 
 \cite{fn3}.  The other  constant $C_\Psi$ is dimensionless and negative.
To complete the model, we note that the real parts are given in terms of  $h(x)$   as
$
\Re e \ \bar{\Phi}(x^+) =  C_\Phi \ h(x) $  and $ \Re e {\Psi}(x^+) =  C_\Psi \ h(x)$. 
With this choice  the auxiliary spectral weight {\em  $\eta(k,x)$  vanishes identically} in  \disp{eta}.
With $\Gamma(x) \equiv \pi C_\Phi \ \sigma(x)$ and $\epsilon(\xi,x) \equiv  \left( x -\xi - C_\Phi \  h(x) \right)$ we may write $ \rho_{\GH}(\xi,x)  = \frac{1}{\pi} \frac{\Gamma(x)}{\Gamma^2(x)+ \epsilon^2(\xi,x)}$  and 
$\Re e \ \GH(\xi, x )= \frac{\epsilon(\xi,x)}{\Gamma^2(x)+ \epsilon^2(\xi,x)}$.
Denoting  $\langle Q(\xi )\rangle_\xi = \int \ d\xi \ N_{B}(\xi) Q(\xi)$, where $N_B(\xi)$ is the band density of states per spin, the chemical potential is fixed  using $\frac{n}{2}= \int_{-\infty}^{\infty} \ dx \  f(x)  \langle\rho_{\GH}(\xi,x)\rangle_\xi$.

The energy parameter  $\Delta(\vec{k},x)$  in \disp{delta} is  a constant. We  scale out a factor to define 
 \barray
\Delta_o = \frac{n^2}{4} \Delta(\vec{k},x)=  -  \frac{n^2}{4} \  \frac{C_\Phi}{C_\Psi}. \label{delta-2}
\earray
 The  physically observable  electronic spectral function reads  
\beq
\rho_{\G}(\xi, x)  =  \frac{\Gamma(x)}{\pi} \frac{  \left(  \left\{ 1- \frac{n}{2}  \right\} + \left(\frac{n^2}{4 }\right) \left\{ \frac{  \xi- x   }{\Delta_0} \right\}    \right)_{+}  }{\Gamma^2(x)+ \epsilon^2(\xi,x)} .
 \label{spectral-3}
\eeq
Here the condition  $(f)_{+}\equiv \mbox{max}(0,f)$,  is   inserted in the ECFL factor to  guarantee the positivity of the spectral function for $x \gg \xi$\cite{fn4}.
 We can determine $\Delta_0$ directly  from the second level  sum rule \disp{higher-order-sumrule}:
\beq
\Delta_0=  \  \int_{-\infty}^\infty  \ dx \ f(x) \ \langle \rho_{\GH}(\xi,x)   \{ \xi- x   \} \rangle_\xi  . \label{delta-eq}
\eeq
Thus $2/n \times \Delta_0$ is the average  inelasticity $||(\xi-x)||$ of the FL Greens function  over the entire  occupied band.  It  vanishes if $\rho_\GH$ were a pure delta function, as in a Fermi gas, but is non zero in a Fermi liquid. 
The linear energy term in \disp{spectral-3} thus  fundamentally arises to  provide the extra density to $\rho_\G$,      compensating    the spectral depletion due to the first factor $1-\frac{n}{2}$ ( originating in the non canonical nature of the projected electrons (\I)). 

\begin{figure}
\includegraphics[width=2.25in]{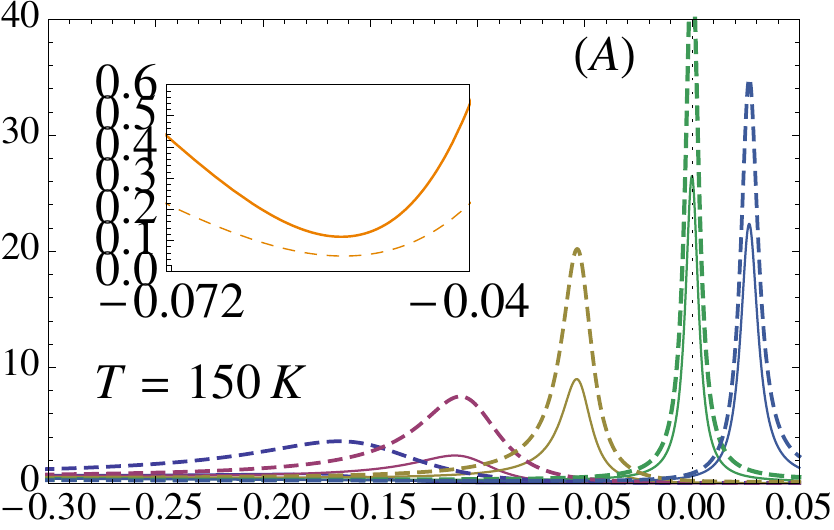}
\includegraphics[width=2.25in]{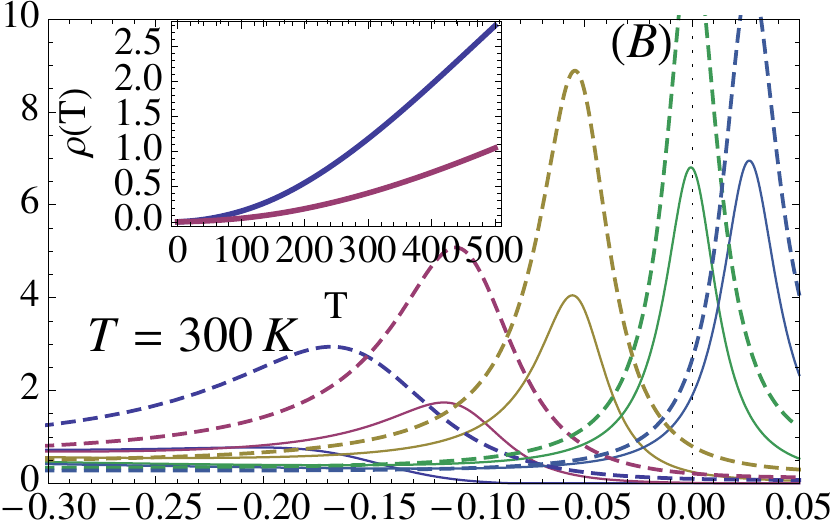}
\includegraphics[width=2.25in]{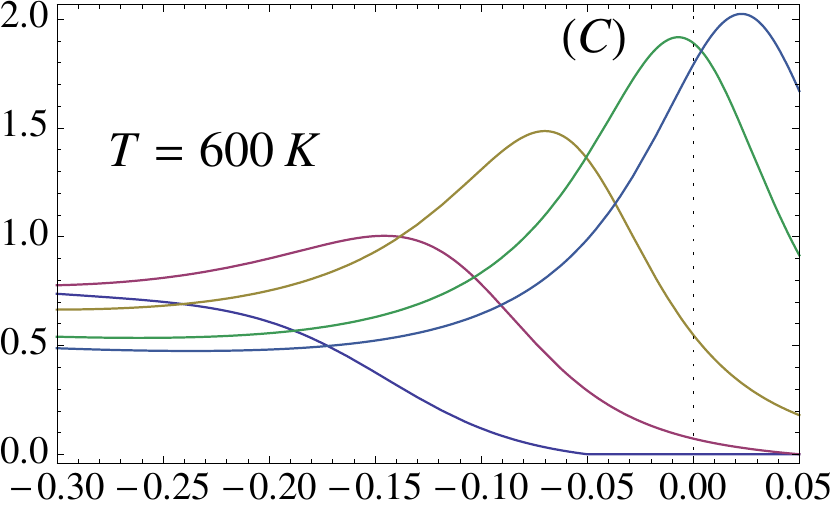}
\caption{ The density $n=.85$ and  $\omega_0=0.25$. From left to right $\rho_{\G}(x)$ for energies (in units of W) $\xi= -0.3,-0.2,-0.1,0.,0.05$ for both the FL (dashed) and the ECFL(solid) theories.  {\bf Inset in (A):}  provides an enlarged view of  the  $\xi=-0.1$  plots after {\em inversion}, and   displays the  left-skew asymmetry of the ECFL spectrum relative to the FL.  {\bf  Inset in (B)} shows the DC resistivity $\rho(T)$ within a bubble approximation as a function of $T$ for the FL (blue) and the ECFL (red). Due to spectral redistribution, the ECFL reaches linear $T$ behaviour at a lower $T$ than the FL.}
\label{Fig_1}
\end{figure}

In the numerical solution of the model, we can vary the shapes of the spectra from sharp to broad by controlling the energy scale $\Delta_0$ via the parameters  $C_\Phi$ and $\omega_0$ in  the FL function $\sigma(x)$.  For illustration we neglect the distinction between the band energy and the renormalized $\eb_k$,   choose a flat band density of states per spin $\rho_0(\varepsilon)= \frac{1}{2 W} \Theta(W^2- \varepsilon^2)$ hence  the  band width  is $2 W$.  Choose $C_\Phi=1$  $W=10^4$K \cite{fn5}, this gives $\Delta_0 \sim 600$K in the cases studied. 
The spectral shapes from \disp{spectral-3} have a characteristic left skew that is visible in \figdisp{Fig_1}, and also in many experimental spectra in high $T_c$ systems. The marginal Fermi liquid hypothesis \cite{mfl}  assumes a linear correction to the spectral function,  but    is {\em symmetric} about the Fermi energy, i.e. of the form $|\xi-x |$ instead of the term in \disp{spectral-3}. 

From \disp{delta-eq}  { a fascinating {\em duality} emerges between the FL and the ECFL}\cite{duality}. When the FL is overall   sharp such that  $\Delta_0$ is small, the ECFL is significantly   broadened. This happens  since in the ECFL factor in  \disp{spectral-3}, the   coefficient of $\xi-x$    becomes large and dominates the $1-\frac{n}{2}$ contribution. 
The function $\Delta(k)$ in \disp{delta} could vanish at points in $k$ space in the full theory (without the assumption of $k$  independence). At those points the ECFL spectra would lose  all coherence by this duality.  A loss of coherence  would  inevitably  suggest a (false)  pseudo gap,  if our current viewpoint  were unavailable.  The linear term also leads to a sloping  term in the local density of states of the ECFL  that the STM technique would  probe, although its magnitude and sign  are less reliably computed- depending as they do on the high energy scales $W$ and $\omega_0$.  In conclusion, we have presented essential ideas underlying    the  theory of  extremely correlated Fermi liquids. We have shown that  an explicit  low order solution is very  promising  in the context of explaining the photoemission spectra of the cuprate materials.

 Detailed numerics and  comparison with experiments are currently underway. This work was supported by DOE under Grant No. FG02-06ER46319.


\begin{thebibliography}{36}
\bibitem{footnote-sym} This expression differs slightly from   \I-29 in \refdisp{ECQL}. The added  terms  vanish exactly  due to the Pauli principle and Gutzwiller projection that eliminate states with double occupancy.  They are added  since they serve to usefully  symmetrize
the expressions for $X(i,j)$ and $Y(i,j)$ and thereby  simplify the subsequent treatment.
\bibitem{fn1} This choice is the essential difference from a decomposition in \I ~ Eq.~(\I-31). In the present case, we are able to establish adiabatic continuity with the Fermi gas as indicated in \disp{EOM-1} below.
\bibitem{lambda} It is important to realize that the nature of the $\lambda$ expansion is different from  that of a typical perturbative expansion, e.g. the $U$ expansion in the Hubbard model. In the present case the scales of  all parameters $t,J,\lambda,\ldots$ are similar  $\sim O(1)$, since the large  parameter $U$ of the Hubbard model has been set at $\infty$ at the outset. Thus $\lambda$ should be viewed as a parameter that organizes the equations in a systematic fashion so that e.g. keeping {\em all terms} of  $O(\lambda^2)$   
gives a consistent theory that is structurally analogous to the skeleton expansion at $O(U^2)$. We may then examine the nature of the solution to this order, with the expectation that the $O(\lambda^3)$ term would retain the qualitative features  found at lower order. Further since   $\lambda$ couples to $(\GH . \mu)^k \sim n$ in \disp{EOM-2},  an  expansion in $\lambda$  should be viewed as a low density expansion.
\bibitem{fn2}  We have thus  imposed the L-W theorem in \disp{higher-order-sumrule} rather than proved it  in this theory.
\bibitem{dispersion-renorm} In \disp{g-1} we  lump all dispersion type terms in the expression  into an effective dispersion   $\eb_k= (1- \lambda \ n ) \varepsilon_k-  \lambda \sum_p E(k,p) \  \GH(p) \ e^{i \omega_p 0^+}$, thus writing $\GHI(k)=  i \omega_k + \chem -  \eb_k  - \lambda \ \bar{\Phi}(k) $.
\bibitem{fn6} This functional  form is based on the $O(U^2)$ result for  of the  Hubbard model, after  one sets the parameter $U\to 1$ in view  of the scales of variables of the \tJ model noted in\cite{lambda}. For small energies the behaviour of the Fermi liquid noted in \cite{agd} is  captured in this form, and a reasonable extrapolation to large frequencies is provided.
\bibitem{fn3} The Hilbert transform of $\sigma(x)$ is denoted by $h(x)$. Here $h(x)=h_1(x)- h_2(x)$ with 
$
h_1(x)=   \pi  (x^2+ \tau ^2) e^{- C_\Phi  \{ x^2+ \tau ^2  \}/\omega_c}  \  \mbox{Erfi}(x \ \sqrt{C_\Phi/\omega_c})$ and $h_2(x)=   \ x \ \sqrt{\pi \omega_c/C_\Phi}  \  e^{- \tau^2 C_\Phi /\omega_c } ,
$
where $  \mbox{Erfi}(x)= \frac{2}{\sqrt{\pi}} \int_0^x e^{t^2} \ dt$ is the imaginary error function.
\bibitem{fn4}In computing the various parameters self consistently, one finds that the positivity enforcing $\Theta$ function in \disp{spectral-3} can be dropped with very little ($\sim 5$\%) error.
\bibitem{fn5} This gives a band width of $\sim 2 $eV that is typical of cuprates.
\bibitem{duality} This physical  duality is to be understood in the sense that  a highly elastic auxiliary FL over all energies, with very sharp features would lead to a small $\Delta_0$ through \disp{delta-eq}, so that the ECFL spectrum would have a very large coefficient of the linear $x$ term in \disp{spectral-3} and hence appear incoherent.
\end{thebibliography}
\end{document}